\documentclass[smallextended]{svjour2}

\newcommand{\xmax}{x_{\footnotesize{\textnormal{max}}}}
\newcommand{\xmin}{x_{\footnotesize{\textnormal{min}}}}

\begin{document}

\title{A Not-So-Fundamental Limitation on Studying Complex Systems with Statistics: Comment on Rabin (2011)}
\author{Drew M. Thomas}
\institute{Drew M. Thomas
           \at Department of Physics, Imperial College London,
               SW7 2AZ, London, UK
           \\\email{dmt107@imperial.ac.uk}}

\journalname{Journal of Statistical Physics}
\date{Received: April 26, 2012 / Accepted: \[TO BE ENTERED\]}

\maketitle

\begin{abstract}
Although living organisms are affected by many interrelated and unidentified variables, this complexity does not automatically impose a fundamental limitation on statistical inference. Nor need one invoke such complexity as an explanation of the ``Truth Wears Off'' or ``decline'' effect; similar ``decline'' effects occur with far simpler systems studied in physics. Selective reporting and publication bias, and scientists' biases in favour of reporting eye-catching results (in general) or conforming to others' results (in physics) better explain this feature of the ``Truth Wears Off'' effect than Rabin's suggested limitation on statistical inference.
\end{abstract}

\vspace{1em}

Yitzhak Rabin's recent paper \cite{Rabin} in this journal draws attention to Jonah Lehrer's article ``The Truth Wears Off'' \cite{Lehrer}, which discusses the apparent fading and shifting over time of experimental results and measurements in science. Rabin outlines a model to try to explain this phenomenon. However, Rabin makes a minor misstatement in the course of discussing his model, and even after allowing for this misstatement his model does not seem to be the best explanation of the ``Truth Wears Off'' effect.

I briefly summarize Rabin's model here to elucidate the misstatement. Consider $N \gg 1$ systems, ``each of which possesses a measurable attribute (variable) $X$'', where $X$ has $M_x \gg 1$ possible values and lies between $\xmin$ and $\xmax$ \cite{Rabin}. An experimenter might measure $X$ for each of the $N$ systems. Rabin writes that for ``a good statistical sample'' one must measure $N \gg M_x$ systems. But this is not generally true, as is shown by considering (like Rabin) the case where $X$ is a continuous variable. Assuming $\xmax > \xmin$, it follows that $X$ can take on infinitely many values and so $M_x$ is infinite. But it does not follow that the sample size $N$ must also be infinite for the sample to be ``a good statistical sample''.

This is so even for a discrete $X$ with finite $M_x$. I find it reasonable to define ``a good statistical sample'' as one where $X$'s empirical cumulative distribution function $G(x)$ differs little from the true cumulative distribution function $F(x)$ of the sampled population. If one measures $X$ in a simple random sample from some population of systems, those $X$ values are independently and identically distributed. Therefore I can immediately apply the Dvoretzky-Kiefer-Wolfowitz-Massart inequality: the probability that $G(x)$ differs anywhere from $F(x)$ by more than some nonzero $\lambda$ is at most $2 \exp(-2 N \lambda^2)$ \cite{Massart}.

Formalizing my definition of ``a good statistical sample'' as one where $|G(x) - F(x)| \le \lambda \ \forall x$, where $\lambda$ is the maximum deviation from the population cumulative distribution function one can tolerate from one's sample, rearranging the inequality gives a probabilistic guarantee that the sample is ``good''. To ensure that the sample is ``good'' with at least probability $p$, the sample size $N$ must satisfy
\begin{equation}
N \ge \frac{\ln \frac{2}{1-p}}{2 \lambda^2}
\end{equation}
When $M_x$ is of similar size to this bound (or greater), $N$ need \emph{not} be much more than $M_x$ for a good sample, in the rather strict sense I formalize it here.

However, this side point --- though worth correcting --- has little bearing on the main argument Rabin makes with his model. That argument is problematic for another reason: Rabin's model does not represent the most common kinds of statistical analyses. Only occasionally do scientists determine whether a variable is interesting or relevant by seeing whether its distribution is flat.

More often, researchers' statistical analyses --- especially in the life sciences --- are based on hypothesis testing, which has a rationale something like the following. One picks a hypothesis to test, makes assumptions about the statistical properties of one's sample, and uses those assumptions to compute a sufficient test statistic $S$ from the sample. If the assumptions hold and the hypothesis is true, $S$ has some probability distribution $p(S)$. But if $S$'s value lies in a low-probability tail of $p(S)$, it becomes implausible that it came from the distribution $p(S)$, and the test rejects the hypothesis.

Does the core of Rabin's qualitative argument carry over to this alternative methodology? At first glance it might seem that it does; $S$ can be influenced by many unknown variables just as $X$ was. But statistical inference furnishes us with a powerful defence against this issue: as long as $S$ comes from the distribution $p(S)$, it does not matter why or how it comes from that distribution; significance tests that use $S$ and $p(S)$ continue to work as advertised.

Consider an opinion poll that asks a few thousand people whether they support a political party. In this scenario one is surely dealing with a complex system affected by many variables: people's views of a political party surely correlate with (and are influenced by) many variables unmeasured by such a poll, such as the political views of one's peers, which television programmes one watches, and so on. As such, Rabin-like arguments by complexity would seem to rule out reliable inferences from this polling technique. But in practice such inferences are possible. How?

It is possible because if one polls a representative, random sample of $N$ people from a large population, then the number of people $S$ in the sample who support a political party is binomially distributed with mean $Np$ and variance $Np(1-p)$, where $p$ is the population proportion of people supporting the party. This is the case however the people in the sample formed their political preferences; it is enough to merely identify a sufficient statistic with a known distribution.

There is an obvious objection to this kind of inference: it relies on an assumption that a test statistic $S$ follows a particular distribution. But this objection can be accommodated by testing the assumption just as if it were any other hypothesis. A researcher can gather many random, representative samples from the same population, calculate $S$'s value for each sample, and then use a test like the $\chi^2$ test or Kolmogorov-Smirnov test to discern whether their sample of $S$ values deviates significantly from $S$'s assumed distribution. In the polling example, one might calculate $S$ values for many polls of sample size $N$ (with the polls carried out at about the same time so that the unknown parameter $p$ changes little across polls), and use a $\chi^2$ test or $G$-test to check that those $S$ values are approximately binomially distributed.

This argument undercuts Rabin's explanation for the difficulty of reproducing certain scientific results. In addition to abstract argument, there is some empirical evidence against Rabin's explanation in the very \textit{New Yorker} article he references \cite{Lehrer}. Rabin's attempted explanation for poor reproducibility points the finger at the complexity of the systems biomedical scientists study, and the large number of mysterious variables that can influence the dependent variables of interest. As such, if Rabin's argument \emph{has} correctly fingered the reason for this effect, the effect should be virtually absent from physicists' studies of parameters describing fundamental particles, which are far simpler systems experimented upon under strongly controlled conditions. Yet Lehrer is able to point to ``the weak coupling ratio exhibited by decaying neutrons, which appears to have fallen by more than ten standard deviations between 1969 and 2001'' (but see \cite{Nichols}) and ``the law of gravity''! More generally, Henrion \& Fischhoff's well-known paper \cite{HF} documents long-standing systematic errors in past estimates of physical constants that were only corrected after much subsequent research.

Hence I disagree on two grounds with Rabin's conclusion that the complexity of complex systems imposes a fundamental limitation on statistical inference. Firstly, in many cases those limitations can be circumvented with the intelligent use of statistics; secondly, less complex systems can be as susceptible to the ``decline effect'' as more complex systems.

I nonetheless agree with Rabin's closing statement that ``critical studies that attempt to test the validity of known results should be encouraged by the scientific community'', subject to mild caveats. (For example, it's surely the case that some results are more in need of testing than others. Cost-benefit analyses could help scientists direct their testing efforts to where they're most needed. It would of course be a waste of resources to have every scientist try to falsify every single result they felt like testing.) But I have more humble reasons for wanting critical scrutiny.
Scientists do have research designs like randomized trials and controlled experiments, and tools like TETRAD \cite{Glymour,Glymour2,Spirtes} that are capable of unpicking causal relationships from a thicket of variables. The problem is not that they can't work but that scientists fail to use them --- and where they \emph{do} use them to perform a critical study, they are less likely to eventually publish the results when they disagree with earlier work.

I infer that much of the blame for the ``Truth Wears Off'' effect should really go to two of the more mundane causes Lehrer suggests: publication bias and selective reporting. This might sound unduly cynical, but I think it's consistent with the evidence. Daniele Fanelli's 2009 meta-analysis of surveys found that, on average, about 10\% of scientists admitted to engaging in any given questionable research practice (QRP) \cite{Fanelli}. A more recent survey of psychologists found that over 90\% ``admitted to having engaged in at least one QRP'' \cite{John}. QRPs often fall in the grey area between best practice and unambiguous misconduct, and can introduce the kind of modest but systematic biases that would lead to the ``Truth Wears Off'' effect. One could read this as a depressing conclusion but I am cautiously optimistic: the problem is less with ``fundamental problems that face any experiments on complex systems'' and more with predictable biases against which we can take precautions.

\end{document}